# Comment on "Macroscopic violation of special relativity" by Nimtz and Stahlhofen [arXiv:0708.0681v1]

Herbert G. Winful

Department of Electrical Engineering and Computer Science, University of Michigan,

1301 Beal Avenue, Ann Arbor, MI 48109-2122

(*Submitted September 18, 2007*)

Abstract

A recent paper by G. Nimtz and A. A. Stahlhofen [arXiv:0708.0681v1] makes the following claims: (1) that the authors have observed a macroscopic violation of special relativity, (2) that they have demonstrated the quantum mechanical behavior of evanescent modes on a meter-length scale, and (3) that barriers are crossed in zero time, implying superluminal (faster than light), and indeed, infinite tunneling velocity. Here I suggest that all these claims are erroneous and are based on a misinterpretation of a purely classical measurement accurately described by Maxwell's equations.

The headline in the August 18, 2007 issue of the *New Scientist* reads "Photons challenge the light barrier; A quantum tunnelling experiment has apparently propelled photons faster than the speed of light." The experiment in question is the one reported in the preprint [2] listed in the title of my Comment and illustrated in Fig. 1. A microwave beam of frequency $f = 9.15$ GHz is incident at an angle of $\theta = 45°$ at the interface between a prism of refractive index $n = 1.6$ and air. Since the incident angle is greater than the critical angle for total internal reflection, the beam is reflected and a small evanescent field penetrates the air gap between the first prism and a second prism placed a distance $d$ from the first. The evanescent field becomes a propagating wave in the second prism and emerges as a transmitted wave. The authors say they have measured the time delay of both the transmitted and reflected beams and find that both beams suffer a delay of 100 ps. It is on the basis of this one result that they make the claim of faster than light propagation and a macroscopic violation of special relativity. Such a claim, if true, would rock the very foundations of modern physics and thus demands investigation.

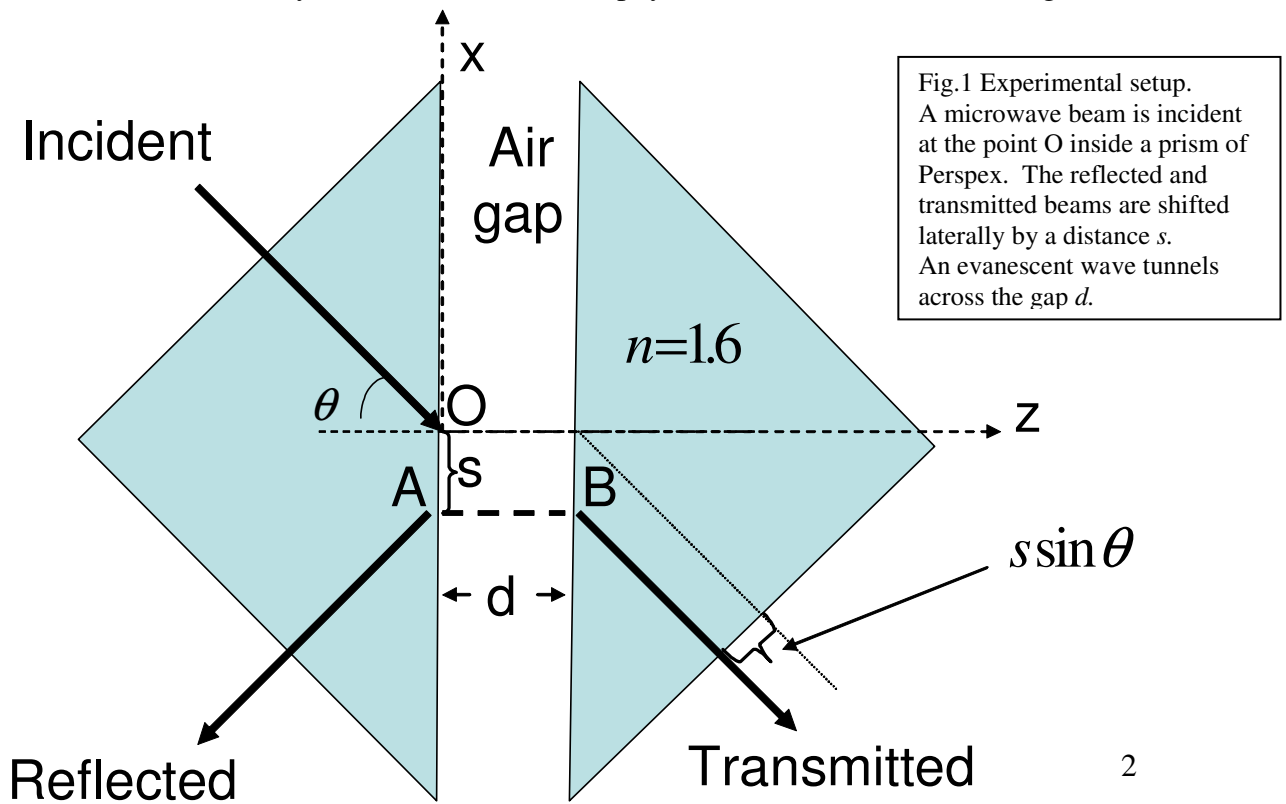

Fig.1 Experimental setup. A microwave beam is incident at the point O inside a prism of Perspex. The reflected and transmitted beams are shifted laterally by a distance *s*. An evanescent wave tunnels across the gap *d*.

The paper itself is short on detail, missing such important information as the width of the pulses used in the experiment and the size of the gap $d$. However, no worries, for it turns out that the experiment that is being reported here was actually published by the Nimtz group as far back as 2000 on the arXiv [3] and then shortly thereafter in several print journals [4-6]. Upon perusing Refs. 3-6 we find that the pulses used had a half width of 8 ns and thus we take the pulse width as $\tau_p = 16$ ns (FWHM). In Refs. 3, 4, and 6 we find plots that show the transmission as a function of gap width for gaps up to 50 mm. Those plots are for microwave carrier frequencies of 8.345 GHz and 9.72 GHz and yield attenuations of 0.73 dB/mm and 0.93 dB/mm, respectively. For the 9.15 GHz frequency used in this reported measurement, we can easily calculate the transmission by using Maxwell's equations and matching boundary conditions at the interfaces. The result for the transmitted field is

$$E_t = E_0 e^{-\kappa d}, \qquad (1)$$

where the attenuation constant is given by

$$\kappa = \frac{2\pi f}{c} \sqrt{n^2 \sin^2(\theta) - 1} \quad \text{m}^{-1}, \qquad (2)$$

and $c = 3 \times 10^8$ ms$^{-1}$ is the speed of light in vacuum. Plugging in the indicated values we find that $\kappa = 101.41$ m$^{-1}$ at 9.15 GHz. The intensity transmission is

$$T = \left| \frac{E_t}{E_0} \right|^2 = e^{-2\kappa d}. \qquad (3)$$

Expressed in dB/mm, the attenuation is

$$10 \log_{10}(e^{-2\kappa \times .001}) = -0.88 \text{ dB/mm}.$$

While the article listed does not actually state the dimension of the air gap, its abstract alludes to "macroscopic scale on the order of a meter." The *New Scientist* quotes the authors as having tunneled photons "instantaneously across a barrier of various sizes, from a few millimeters up to a meter." We can believe a few millimeters. But at 0.88 dB/mm, an evanescent wave tunneling across a 1 meter gap would be attenuated by 880 dB! This translates to a transmission of $10^{-88}$. Suffice it to say that the authors could not possibly have measured tunneling across a 1 m gap, a result that has set internet discussion groups all aquiver. The maximum spacing they could have used and still obtain a measurable transmission would be about 40 mm, which would yield an attenuation of 35.2 dB, consistent with the minimum transmissions reported in Refs. 3, 4, and 6.

A closer reading of the paper under Comment reveals the statement "The sides of the right triangle prisms are $0.4 \times 0.4$ m$^2$, which is of a macroscopic dimension for a quantum mechanical experiment." This statement, while technically true, is highly misleading. First, there is nothing quantum mechanical about this experiment. Everything about it can be predicted purely on the basis of the classical Maxwell equations. The transmission as calculated classically agrees perfectly with the experimental results as the authors themselves note in Refs. 3-6. Second, the dimensions of the sides of the prisms are completely irrelevant as far as tunneling is concerned. The only relevant dimension is the gap *d* between the prisms and that is at most a few centimeters.

The claim of a violation of special relativity is one that cannot be taken seriously. The experiment as noted is completely described by Maxwell's equations which are Lorentz invariant and therefore strictly in accord with special relativity. The measured delay times agree with the predictions based on Maxwell's equations [7] and so the validity of special relativity is actually *confirmed* by this experiment. It is curious to note that in a previous paper on tunneling through undersized waveguides [8], Nimtz himself stated: "Therefore microwave tunneling, i.e. the propagation of guided evanescent modes, can be described to an extremely high degree of accuracy by a theory based on Maxwell's equations."

What then is this 100 ps delay that was measured to be equal for both reflected and transmitted pulses? Is there any implication of superluminal propagation? Let us take a closer look at the experiment and the interpretation put forth by Nimtz and Stahlhofen (N&S). Consider a pulse incident at the point marked O (Fig. 1) in the first prism at $t = 0$. The reflection group delay is defined as the time at which the reflected pulse peak appears at A. The transmission group delay is the time at which the peak of the transmitted pulse appears at B. Both the reflected and transmitted waves undergo a lateral shift known as the Goos-Hänchen shift, a distance indicated by *s* in Fig. 1. In addition, Nimtz and Stahlhofen *assume*, and we emphasize that this is only an assumption, that inside the gap the tunneling pulse follows the path AB of length *d*. Based on this assumption, the authors conclude that the tunneling pulse traveled a distance *d* + *s* in the same time that the reflected pulse covered the distance *s* and hence the tunneling wave must have covered the extra distance *d* in zero time. The assumption that the tunneling evanescent wave follows a definite path AB in the gap has no basis. Inside the gap the evanescent wave has an imaginary wave vector in the direction $\overrightarrow{AB}$ and hence one cannot assign it a path as if it were some point particle moving in a straight line. Within the gap the evanescent wave exists as an exponential standing wave that fills the entire gap (see Fig. 2).

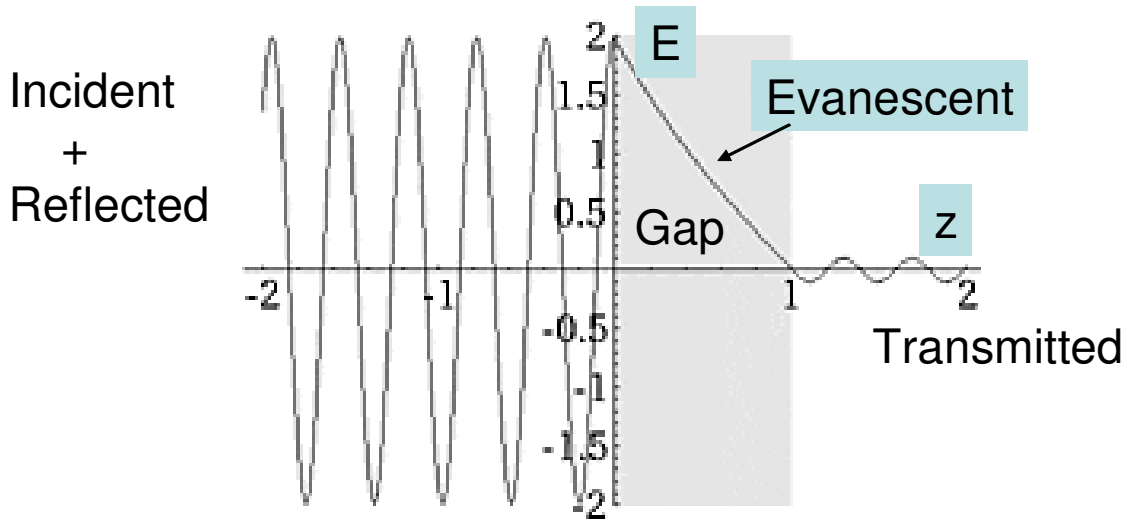

Fig. 2 Nature of the evanescent wave in the air gap. It is an exponentially decaying field that simply moves up and down in time. In the first prism is a standing wave made up of the interference between incident and reflected waves. The transmitted field is a traveling wave.

A theoretical calculation of the group delay $\tau_g$ based on Maxwell's classical electrodynamics yields two terms [9,10]:

$$\tau_g = \tau_0 + s\left(\frac{n\sin\theta}{c}\right). \qquad (4)$$

The first term is related to a derivative of the transmission (or reflection) phase shift with respect to angular frequency. According to theory, the group delay is the same for both reflected and transmitted pulses. Now Nimtz and Stahlhofen wrongly associate the first term $\tau_0$ with a transit time of a pulse from A to B. That term goes to zero rapidly for gaps of the order of a wavelength or larger. Because that term goes to zero, N & S jump to the conclusion that a pulse went from A to B instantaneously, with infinite velocity, smashing the speed of light. There is absolutely no justification for that conclusion.
Inside the gap the electric field is of the form

$$E = (Ce^{-\kappa z} + De^{\kappa z})e^{i(k_x x - \omega t)},$$

where $C$ and $D$ are arbitrary constants determined by the boundary conditions and $k_x = (n\omega/c)\cos\theta$. This is an inhomogeneous plane wave that propagates along $x$ (parallel to the interfaces) with an amplitude that is evanescent (decays) or anti-evanescent (grows) with $z$. For wide enough barriers we can drop the anti-evanescent term with amplitude $e^{+\kappa z}$ since it blows up at infinity. What we have is a standing wave in the $z$ direction. In a standing wave, the field at all points, and in particular at A and B, moves up and down in phase. It doesn't mean that anything has traveled from A to B in zero time. We stress this point because that is the whole basis for the claim by Nimtz and Stahlhofen that they have overturned special relativity. In the gap there is no real propagation in the $z$ direction and hence $\tau_0$ is not a transit time from A to B.

The second term in the group delay expression is the dominant one for thick barriers. It is proportional to the Goos-Hänchen shift $s$. Because there is true propagation in the $x$ direction, we can associate this term with the time it takes for energy to be transported a distance $s$ in a direction parallel to the interfaces. The overall group delay $\tau_g$ thus represents the lifetime of stored energy in the region bounded by the distances $d$ along $z$ and $s$ along $x$ (per unit depth along $y$). This lifetime is proportional to the energy stored. It turns out that it is much easier to measure a lateral beam shift than to directly measure a short time delay. N&S state that their receiver antenna "was movable parallel to the prism's surfaces." By moving this antenna they are able to measure $s\sin\theta$ which can then be used in Eq. 4 to determine the group delay, keeping in mind that the first term is vanishingly small. Indeed, if we work backward and use 100 ps for the second term of the group delay expression, we find $s = 2.65$ cm, in agreement with the shifts reported in Ref. 5 for wide beams (see inset of Fig.3 in Ref. 5). Other authors have used similar measurements of the Goos-Hänchen shift to suggest that tunneling beams travel with superluminal group velocity [11]. In that regard the claim in the Nimtz and Stahlhofen article is not new. Of course the group delay can also be determined directly from two time delay measurements: (1) the delay $t_1$ between the generated pulse at the transmitter and the received pulse at the horn antenna with the prisms in the closed position and (2) the delay $t_2$ between source and receiver with the prisms separated by the gap $d$. The difference $t_1 - t_2$ yields the group delay due to the gap. It is surprising that there is no actual data on this direct determination of group delay anywhere in the Nimtz papers referenced – no figures showing received pulses with the prisms closed and with the prisms apart. Other workers have shown such pulse data from FTIR experiments because they tell the whole story [12]. Here the only hard data presented in support of this 100 ps group delay are the measured Goos-Hänchen shifts for the *reflected* beam reported in 2001 in Ref. 5. It is upon such data that N&S wish to overturn special relativity. In all their published oeuvre we have not found a single figure or data point that shows either the Goos-Hänchen shift or the pulse temporal delay for the *transmitted* (tunneled) beam in frustrated total internal reflection. (Of course, it's much easier to monitor the reflected beam since it contains 99.99% of the incident energy.)

To put everything into perspective, one must look at the relative sizes of things. Here we list the important quantities involved:

1) Temporal pulse width: $\tau_p = 16$ ns   Spatial extent of pulse: $c\tau_p = 4.8$ m .

2) Wavelength = 3.28 cm.

3) Group delay =100 ps.   Spatial extent of delay=3 cm.

4) Gap width = 4 cm.

5) Transmission for 4 cm gap = $3 \times 10^{-4}$

The first thing to note is that the spatial extent of the pulse, 4.8 m, is two orders of magnitude greater than the width of the gap (4 cm). Figure 3 illustrates the relative dimensions of pulse and gap. The pulse is so long that it cannot be localized within the gap. One cannot say that at $t = 0$ the pulse is definitely at A and not at B and that at $t = \tau_g$ it has crossed to B and is no longer at A. In fact this is a quasi-static interaction [13]: the pulse is so long that at every instant the system is pretty much in steady state. The barrier is a lumped element with respect to the envelope and acts just like a capacitor that can store energy. For such a lumped system the concept of a "group velocity" for traversing the gap is meaningless while group *delay* remains a perfectly valid concept describing how long the capacitor stores energy before it decays away.

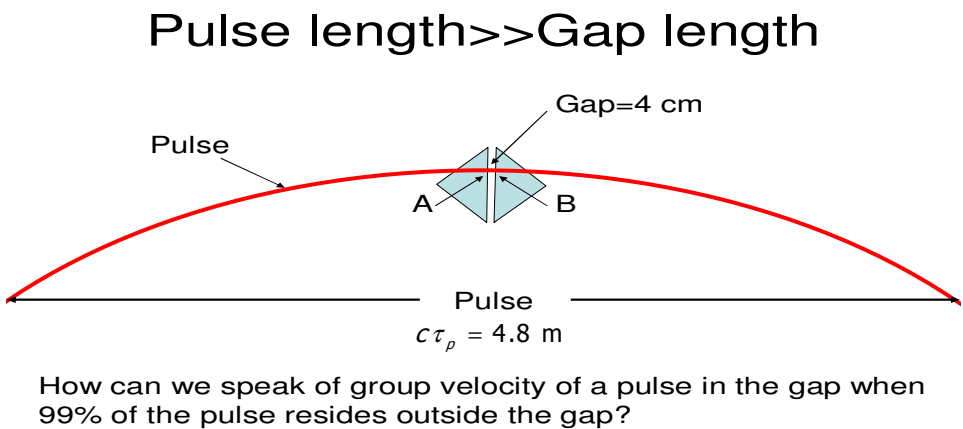

Fig. 3. Relative dimensions of the microwave pulse and the gap between the prisms. The pulse is much longer than the barrier which acts as a lumped element (like a capacitor).

The second thing to note is that the delay of 100 ps is a small fraction of the 16,000 ps pulse width. This delay is not a traversal time of anything traveling from A to B across the gap d. It is simply the time it takes for energy stored in the gap, within the volume of cross section $s \times d$, to escape through the reflection, transmission, and lateral shift channels [14]. This time is equal for both reflected and transmitted pulses as one would expect of stored energy leaking out of both sides of a symmetric cavity (see Fig. 4). This cavity lifetime is proportional to the stored energy. The stored energy in the evanescent wave is less than would be stored in a freely propagating wave occupying the same volume. That is why the delay is shorter than for the free wave. It should also be noted that the strength of the transmitted pulse is about four orders of magnitude lower than the incident pulse, which means that the tunneled intensity, even at its maximum is always less than the intensity of the free-traveling pulse.

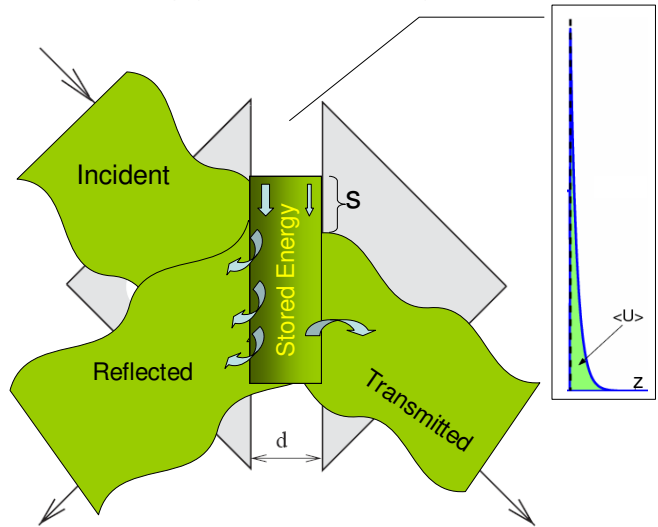

Fig. 4 Energy is stored briefly in an exponentially decreasing evanescent wave. The energy is transported vertically (downwards) to produce the Goos-Hanchen shift while leaking through the sides. Most of the escaping energy goes into the reflected wave. Inset: the exponential decay of energy density $\langle U \rangle$ with $z$.

According to the *New Scientist*, Nimtz and Stahlhofen also found that "the tunneling time, if any, did not change as they pulled the prisms further apart." This would be difficult to explain if the "tunneling time" were a transit time from A to B, for how would the pulse be so smart that it knows the distance has been increased and so it should speed up to cover the larger distance in the same time? On the other hand, this effect, known as the Hartman effect [15], is easy to explain if the delay is the lifetime of stored energy [16,17]. The energy density in the wave decays exponentially with distance. The total energy is the integral of this energy density. This total energy saturates with gap length. Since the group delay is proportional to the stored energy, it saturates in the same manner with gap length. The saturation of the group delay and the saturation of the Goos-Hänchen shift are linked since it is the shift that determines the delay. A physical picture of the shift and delay is that of an incident beam penetrating and filling the gap and then moving vertically a distance s before spilling out of both sides of the gap. Under quasi-steady state conditions, the stored energy in the barrier flows out continually from both sides and must be replenished by the incoming flux. This is what leads to a finite delay between incident and transmitted peaks.

Einstein liked to use trains to explain aspects of special relativity. Here, to demonstrate that his theory is safe, we will use a train analogy to explain the Hartman effect. First suppose that the group delay measures the time it takes to empty a train of all the passengers on board. Clearly the more people on the train the longer it will take to empty the train. Suppose the cars in the train are occupied in the following manner: each successive car holds only one-half the number of passengers in the preceding car.  In other words the cars are loaded according to the formula N, N/2, N/4, N/8, N/16,… Then, no matter how long we make the train, the number of passengers on board will never exceed 2N.  Since the delay is proportional to the number of passengers on the train, the delay will also saturate with the length of the train.

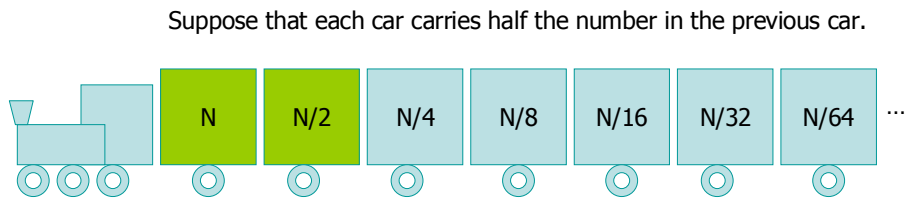

Fig. 5  The train analogy for the Hartman effect.

The *New Scientist* article also relates another train analogy that was used to try to explain the results of Nimtz and Stahlhofen.  A 20-car train leaves Chicago for New York.  "The stopwatch starts when the centre of the train leaves the station, but the train leaves cars behind at each stop. So when the train arrives in New York, now comprising only two cars, its centre has moved ahead, although the train itself hasn't exceeded its reported speed."  Note that according to this analogy the train that arrives in New York is much shorter than the one that left Chicago.  This analogy is a variant of the so-called

"reshaping argument" for the existence of superluminal tunneling velocities [18]. According to that argument, the barrier transmits the front end of the pulse and chops off the back end, resulting in a forward shift of the pulse's peak and a shortening of the pulse. Unfortunately this argument is supported neither by the experimental observations [19,20] nor by simulations [21]. In all cases the transmitted pulse is the same length and the same shape as the incident pulse, albeit much attenuated in intensity. The reshaping argument simply does not apply to tunneling pulses and needs to be laid to rest.

We conclude by summarizing our main points:

1) The experiment of Nimtz and Stahlhofen is a purely classical one accurately described by Maxwell's equations. It is not a quantum mechanical experiment.

2) Because it is described by the fully relativistic Maxwell's equations, the results do not violate special relativity. Rather, they affirm the special theory of relativity.

3) Nothing was observed to be traveling faster than light. The measured delay is the lifetime of stored energy leaking out of both sides of the barrier. The equality of transmission and reflection delays is what one expects for energy leaking out of both sides of a symmetric barrier.

4) The authors could not possibly have observed tunneling across a meter gap since the transmission would have been -880 dB or $10^{-88}$.

5) The saturation of delay with gap length (Hartman effect) is due to the saturation of stored energy.

6) You do not need virtual particles to understand these effects.

**REFERENCES**

1. New Scientist, August 18, 2007, page 10.

2. G. Nimtz and A. A. Stahlhofen, arXiv:0708.0681v1 (2007).

3. A. Haibel and G. Nimtz, arXiv:physics/0009044v1 (2000).

4. A. Haibel and G. Nimtz, Ann. Phys. (Leipzig), **10**, 707 (2001).

5. A. Haibel, G. Nimtz, and A. A. Stahlhofen, Phys. Rev. E **63**, 047601 (2001).

6. G. Nimtz, Prog. in Quantum Electron. **27**, 417 (2003).

7. A. Ghatak and S. Banerjee, Appl. Opt. **28**, 1960 (1989).

8. H. M. Brodowsky, W. Heitmann, and G. Nimtz, Phys. Lett. A **222**, 125 (1996).

9. A. M. Steinberg and R. Y. Chiao, Phys. Rev. A **49,** 3283 (1994).

10. C.-F. Li, Phys. Rev. A **65**, 066101 (2002).

11. P. Balcou and L. Dutriaux, Phys. Rev. Lett. **78,** 851 (1997).

12. M.T. Reiten, D. Grischkowsky, and R.A. Cheville, Phys. Rev. E **64,** 036604 (2001).

13. H. G. Winful, Phys. Rev. Lett. **90,** 023901 (2003).

14. H. G. Winful, New J. Phys. **8,** 101 (2006).

15. T. E. Hartman, J. Appl. Phys. **33**, 3427 (1962).

16. H. G. Winful, Opt. Express **10**, 1491 (2002).

    http://www.opticsinfobase.org/abstract.cfm?URI=oe-10-25-1491

17. H. G. Winful, IEEE J. Sel. Top. In Quantum Electron. **9,** 17 (2003).

18. M. Büttiker and S. Washburn, Nature **422,** 271 (2003).

19. S. Longhi, M. Marano, P. Laporta, and M. Belmonte, Phys. Rev. E**64,** 055602 (2001).

20. S. Doiron, A. Hache, and H. G. Winful, Phys. Rev. A **76,** 023823 (2007).

21. H. G. Winful, Nature (London) **424,** 638 (2003).